\documentclass[aps, pra, reprint, amsmath, amssymb, superscriptaddress, showpacs]{revtex4-1}
\usepackage{CJK}
\usepackage{graphicx}
\usepackage{braket}
\usepackage{mathtools}

\usepackage[usenames,dvipsnames]{color}
\usepackage{hyperref}
\hypersetup{colorlinks,citecolor=blue,linkcolor=blue,urlcolor=blue}
\usepackage{soul}

\begin{document}

\title{Dynamics of Fluctuations in a Quantum System}

\author{Yi-Jen Chen}
\email{yi-jen.chen@cfel.de}
\affiliation{Center for Free-Electron Laser Science, DESY, Notkestrasse 85, 22607 Hamburg, Germany}
\affiliation{Department of Physics, University of Hamburg, Jungiusstrasse 9, 20355 Hamburg, Germany}

\author{Stefan Pabst}
\affiliation{Center for Free-Electron Laser Science, DESY, Notkestrasse 85, 22607 Hamburg, Germany}

\author{Zheng Li}
\affiliation{Center for Free-Electron Laser Science, DESY, Notkestrasse 85, 22607 Hamburg, Germany}
\affiliation{Department of Physics, University of Hamburg, Jungiusstrasse 9, 20355 Hamburg, Germany}

\author{Oriol Vendrell}
\affiliation{Center for Free-Electron Laser Science, DESY, Notkestrasse 85, 22607 Hamburg, Germany}

\author{Robin Santra}
\email{robin.santra@cfel.de}
\affiliation{Center for Free-Electron Laser Science, DESY, Notkestrasse 85, 22607 Hamburg, Germany}
\affiliation{Department of Physics, University of Hamburg, Jungiusstrasse 9, 20355 Hamburg, Germany}

\date{\today}

\begin{abstract}
``\textit{The noise is the signal}''[R.~Landauer, Nature \textbf{392}, 658 (1998)] emphasizes the rich information content encoded in fluctuations. This paper assesses the dynamical role of fluctuations of a quantum system driven far from equilibrium, with laser-aligned molecules as a physical realization. Time evolutions of the expectation value and the uncertainty of a standard observable are computed quantum mechanically and classically. We demonstrate the intricate dynamics of the uncertainty that are strikingly independent of those of the expectation value, and their exceptional sensitivity to quantum properties of the system. In general, detecting the time evolution of the fluctuations of a given observable provides information on the dynamics of correlations in a quantum system.
\end{abstract}


\maketitle

For a general quantum system, a measurement of an observable is, in principle, of probabilistic nature. For many macroscopic thermodynamic quantities of a large system in equilibrium, the ensemble average is sufficient to represent the outcome distribution, and fluctuations from the average, quantitatively characterized by the uncertainty, are negligible \cite{d.chandler87a}. Nevertheless, there has been a growing number of studies evidencing that fluctuations can play a key role in various systems, with recent impetus from advanced experimental techniques that enable the manipulation of small systems or the resolution of microscopic non-coarse-grained properties to an unprecedented degree. For instance, fluctuations of persistent currents in mesoscopic metal rings \cite{m.buttiker83a, e.riedel93a, v.chandrasekhar91a} shed light on the electronic static correlation and diffusion. This long-standing controversy has been solved lately using a scanning superconducting quantum interference device \cite{h.bluhm09a} and microtorsional magnetometry \cite{a.bleszynski-jayich09a}. Also, density fluctuations of a Bose-Einstein condensate (BEC) in an optical lattice manifest the quantum phase transition to a Mott insulator \cite{d.jaksch98a, c.orzel01a, m.greiner02a}, which has been directly probed by means of a newly developed single-atom- and single-site-resolved fluorescence imaging technique \cite{j.sherson10a}.

The aforementioned examples illustrate the importance of static fluctuations in an equilibrium system. Now it is even more intriguing to consider the dynamics of fluctuations and their information content in a nonstationary quantum system. To the best of our knowledge, such examples are scarce. A recent work on nuclear magnetic resonance in a semiconductor nanowire perturbs and monitors the real-time spin fluctuations in demonstration of their ability to harness spin noise \cite{p.peddibhotla13a}. Another experiment quenches a spin-1 BEC in an optical trap and measures the fluctuations of the freely evolved spin population, which indicate the limitations of mean-field theory \cite{c.gerving12a}. In these specific cases, the temporal variation in the uncertainty predominantly follows the temporal variation in the expectation value in a monotonic fashion. That is, the dynamical behavior of the uncertainty can be intuitively inferred from that of the expectation value.

In this paper, we assess the dynamics of the fluctuations of an observable that are not intuitively related to the dynamics of the expectation value of that observable in a quantum system driven far from equilibrium. For this purpose, an ensemble of laser-aligned molecules \cite{h.stapelfeldt03a, t.seideman06a} is chosen as a physical realization, a system itself of ongoing attention across various disciplines such as molecular strong-field and attosecond physics \cite{t.kanai05a, m.mechel08a} and diffractive imaging of isolated molecules with x rays \cite{s.pabst10a, j.kupper14a} or electron pulses \cite{c.hensley12a}. We study the time evolution of the expectation value of a commonly used one-body operator $\langle \hat{O} \rangle$ as well as that of the uncertainty $\Delta O \equiv ( \langle \hat{O}^{2} \rangle - \langle \hat{O} \rangle ^{2} )^{1/2}$ by numerically solving the exact time-dependent Schr\"{o}dinger equation of the model system and its classical counterpart. In contrast to the above spinor systems \cite{p.peddibhotla13a, c.gerving12a}, the laser-driven rotors provide insight that the dynamics of the uncertainty can be completely decoupled from those of the average: The uncertainty displays significant and intricate motion while the average is temporarily static. The counterintuitive dynamics of the uncertainty reflect higher order quantum coherence in the molecular density matrix and a hidden order of the alignment distribution, both inaccessible through the dynamics of the average. Also, we show that the time evolutions of the uncertainty more efficiently distinguish between quantum and classical descriptions. Finally, we point out that the dynamics of fluctuations of a one-body observable may be a general tool for gaining insight into the dynamics of two-body correlations in a nonequilibrium quantum many-body system.

For computational implementation, we consider a thermal ensemble of gas-phase bromine molecules $^{79}\text{Br}-^{81}\text{Br}$ driven by an intense nonresonant laser pulse linearly polarized along the $z$ axis. Molecular relaxation via collisions is not included, since it does not much affect the alignment dynamics on the time scale of interest in normal molecular beam experiments \cite{p.ho09a, t.seideman06a}. The total Hamiltonian for each molecule is $\hat{H}(t) = \hat{H}_{\text{rot}} + \hat{H}_{\text{int}}(t)$. Here, $\hat{H}_{\text{rot}} = B \hat{\mathbf{J}}^{2}$ represents the Hamiltonian for a free rigid rotor, with $B$ the rotational constant and $\hat{\mathbf{J}}$ the angular momentum operator. $\hat{H}_{\text{int}}(t)$ describes the laser-molecule interaction \cite{p.ho09a, c.buth08a, t.seideman06a}:
\begin{equation}
\hat{H}_{\text{int}}(t) = -2\pi \alpha I(t) \Delta \alpha \left( \cos ^{2} \theta - \frac{1}{3} \right), \label{eq:int_hamiltonian}
\end{equation}
where $\alpha$ is the fine-structure constant, $I(t)$ the cycle-averaged pulse intensity, $\Delta \alpha$ the dipole polarizability anisotropy, and $\theta$ the polar angle between the molecular axis and the laser polarization axis. The conventional observable adopted in the literature is $\hat{O} = \cos ^{2} \theta$ \cite{h.stapelfeldt03a, t.seideman06a}, an order parameter specifying the degree of alignment, and typically only its expectation value is of concern \cite{h.stapelfeldt03a, j.ortigoso99a, k.lee06a, a.goban08a, j.cryan09a, d.pentlehner13a}. We solve the quantum dynamics of the expectation value $\langle \cos ^{2} \theta \rangle$ and the uncertainty $\Delta \cos^{2}\theta$ using the \textit{ab initio} numerical techniques detailed in Refs.~\cite{c.buth08a, s.pabst10a}. Since uncertainty lies at the heart of quantum mechanics, the classical dynamics are also calculated with the recipe from Ref.~\cite{p.ho09a}. This allows us to scrutinize the quantum-classical correspondence not only for the average \cite{p.ho09a, m.leibscher03a, y.khodorkovsky11a, j.hartmann12a, m.molodenskii07a} but also for the uncertainty. The molecules are initially at a rotational temperature $0.5 \: \text{K}$ and are pumped by a Gaussian pulse centered at $t = 0 \: \text{ps}$. The pulse's peak intensity $I_{\text{max}}$ is tuned to obtain a maximum degree of alignment of $\langle \cos ^{2} \theta \rangle \approx 0.8$. By varying the FWHM pulse duration $\tau_{\text{FWHM}}$, we discuss the time evolutions of the expectation value and the uncertainty in the two most instructive cases: the adiabatic and the impulsive limits.

\begin{figure}[t]
\includegraphics[width=86mm]{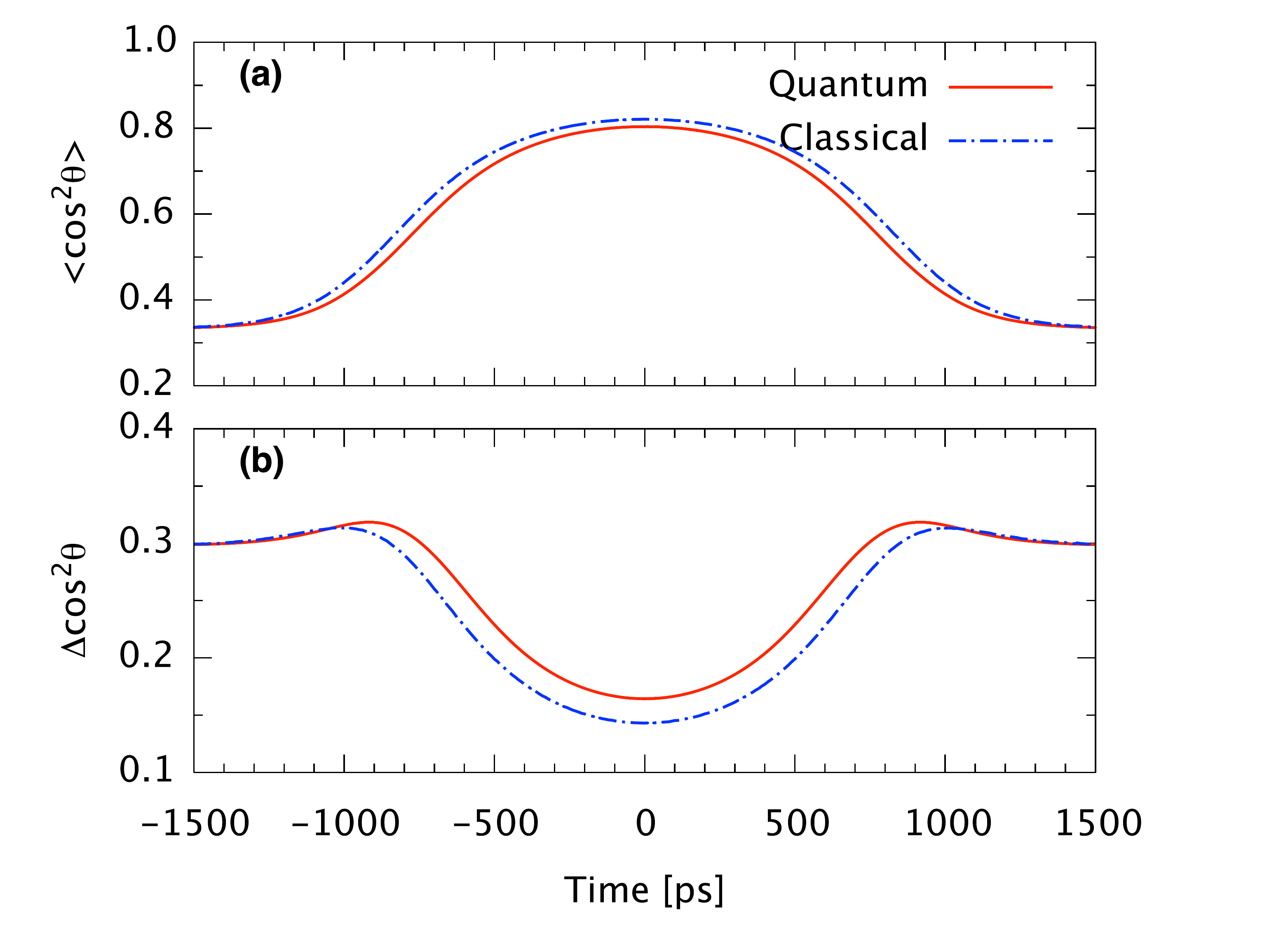}
\caption{\label{fig:adiabatic}(Color online) Laser-induced alignment dynamics of $\text{Br}_{2}$ molecules under adiabatic conditions: (a) mean $\langle \cos^{2}\theta \rangle(t)$ and (b) fluctuation $\Delta \cos^{2}\theta(t)$. Red solid curves for quantum results; blue dot-dashed curves for classical ones. See text for parameters used.}
\end{figure}

Figure \ref{fig:adiabatic} illustrates the dynamics of $\langle \cos^{2}\theta \rangle$ and $\Delta \cos^{2}\theta$ in the adiabatic regime, where the pulse is slowly ramped with respect to the characteristic molecular rotational period of $420 \: \text{ps}$. The parameters used are $\tau_{\text{FWHM}} = 1 \: \text{ns}$ and $I_{\text{max}} = 2 \times 10^{11}\: \text{W}/\text{cm}^{2}$. First we focus on the quantum dynamics. The quantum mean value $\langle \cos ^{2} \theta \rangle$ in panel (a) is a smooth curve mimicking $I(t)$. The alignment mechanism in this scenario is described by the adiabatic evolution of each field-free rotational eigenstate into an instantaneous field-dressed pendular state \cite{h.stapelfeldt03a, b.friedrich95a}. The quantum fluctuation $\Delta \cos ^{2} \theta$ in panel (b) is mainly a function anticorrelated with $\langle \cos ^{2} \theta \rangle$. For $\cos ^{2} \theta$ probes microscopic properties of individual molecules, the magnitude of $\Delta \cos ^{2} \theta$ is non-negligible. However, the dynamical behavior of the uncertainty is not truly surprising: We expect a reduction of the uncertainty when the average approaches its extreme value(s). In our case, this means $\Delta \cos ^{2} \theta \rightarrow 0$ as $\langle \cos ^{2} \theta \rangle \rightarrow 0, \:1$. Such is the intuitive dynamical relation between expectation values and uncertainties that can also be found in Refs.~\cite{p.peddibhotla13a, c.gerving12a}.

Next we compare the classical and quantum dynamics in Fig.~\ref{fig:adiabatic}. Albeit adiabatic alignment is usually considered an intensively studied problem readily understood in classical terms \cite{h.stapelfeldt03a}, the classical and quantum averages and uncertainties still show quantitative discrepancies. The subtle quantum effect is especially apparent for $\Delta \cos^{2}\theta$. Interestingly, the classical approach always overestimates $\langle \cos^{2}\theta \rangle$ yet underestimates $\Delta \cos^{2}\theta$, so classical molecules are easier to align. This is because the pendular states slightly penetrate into the angular potential barrier \cite{w.kim98a}, while the classical molecules are strictly confined to move between the classical turning points.

\begin{figure}[t]
\includegraphics[width=86mm]{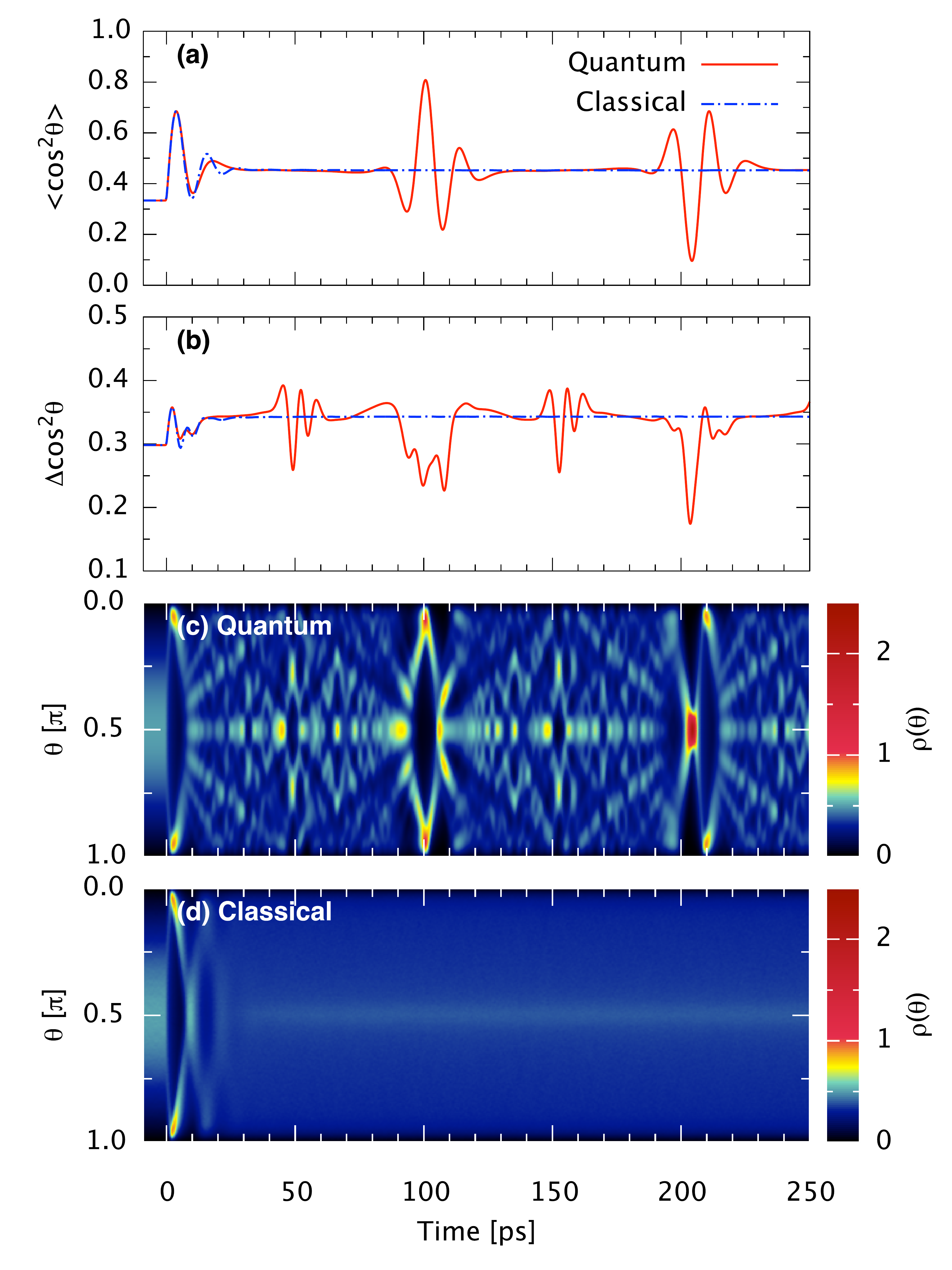}
\caption{\label{fig:impulsive}(Color online) Laser-induced alignment dynamics of $\text{Br}_{2}$ molecules under impulsive conditions: (a) mean $\langle \cos^{2}\theta \rangle(t)$ and (b) fluctuation $\Delta \cos^{2}\theta(t)$. In above-mentioned panels, red solid curves for quantum results; blue dot-dashed curves for classical ones. For comparison, time evolutions of total angular probability distributions $\rho(\theta)$ computed quantum mechanically and classically are plotted in (c) and (d), respectively. See text for parameters used.}
\end{figure}

Figure \ref{fig:impulsive} depicts the alignment dynamics in another limit, the impulsive regime, where a short pulse is used with $\tau_{\text{FWHM}} = 500 \: \text{fs} $ and $I_{\text{max}} = 2.0 \times 10^{12}\: \text{W}/\text{cm}^{2}$. The quantum mean $\langle \cos^{2}\theta \rangle$ reaches the first peak slightly after $t=0$, followed by transient recurrences of the signal and an upward baseline shift. The initial alignment is fully reconstructed at the first rotational revival $\tau_{\text{rev}} = \pi \hbar/B \approx 210 \: \text{ps}$ (with $\hbar$ the reduced Planck constant), and is inverted at $1/2 \:\tau_{\text{rev}}$. The mechanism here is that the kick pulse creates an initially localized rotational wave packet, whose field-free components continue to dephase and rephase periodically owing to their commensurate energy spacings \cite{h.stapelfeldt03a, t.seideman99a}.

The quantum fluctuation $\Delta \cos ^{2} \theta$ exhibits a similar elevated baseline but wiggles dramatically at every quarter revival. At $1/2 \:\tau_{\text{rev}}$ and $\tau_{\text{rev}}$, $\Delta \cos ^{2} \theta$ is again (anti)correlated with $\langle \cos^{2}\theta \rangle$. Nonetheless, at $1/4 \: \tau_{\text{rev}}$ and $3/4 \: \tau_{\text{rev}}$, where $\langle \cos^{2}\theta \rangle$ is essentially static and suggests that the alignment order is temporarily lost due to wave-packet dephasing, $\Delta \cos ^{2} \theta$ displays contrasting and independent beatings of a magnitude comparable to $\langle \cos^{2}\theta \rangle$. This demonstrates that the uncertainty can have nontrivial and intricate dynamics notably distinct from those of the average alignment order parameter and that, in this case, both of them are equally crucial for understanding the system's dynamical properties.

The origin of the additional beatings in $\Delta \cos ^{2} \theta$ can be explained based on an extended argument for the dynamical behavior of $\langle \cos^{2}\theta \rangle$ \cite{s.ramakrishna05a, s.fleischer12a}. Consider the molecular density operator in the representation of the field-free rotational eigenstates $\ket{J,M}$ (with quantum numbers $J$ for total and $M$ for the $z$component of angular momentum): $\Bra{JM}\hat{\mathbf{\rho}}(t)\Ket{J'M'}$. From the underlying symmetries, $\langle \cos^{4} \theta \rangle = \text{Tr}[\hat{\mathbf{\rho}}(t)\cos^{4}\theta]$ involved in the evaluation of $\Delta \cos ^{2} \theta$ can be partitioned into three components summing over different $\Delta J\equiv (J-J')=0,\: \pm 2, \: \pm4$ but the same $\Delta M \equiv(M-M')= 0$. The first term with $\Delta J = 0$ measures the diagonal density matrix elements, i.e., the populations redistributed by the pulse, resulting in a constant background shift. The second part with $\Delta J = \pm2$ probes the off-diagonal elements, i.e., the induced quantum coherences. Because the energy splitting of this type of coherence pairs is $2B(2J+3)$, this part contributes to a signal in $\langle \cos^{4} \theta \rangle$ beating with a period $\tau_{\text{rev}}$. Similarly, the last term also extracts off-diagonal elements but a different coupling pair with $\Delta J = \pm4$. Following an energy splitting of $4B(2J+5)$, this type of coherences leads to a signal reviving every $1/2\:\tau_{\text{rev}}$ and with a flipped sign at odd multiples of $1/4\:\tau_{\text{rev}}$. Consequently, the dynamics of $\Delta \cos ^{2} \theta$ manifest the additional quantum coherences $\Delta J = \pm4$ in the density matrix, whereas the dynamics of $\langle \cos^{2}\theta \rangle$ reflects coherences only up to $\Delta J = \pm2$. As subgroups of wave packet components partially rephasing at fractional revivals is a general phenomenon for an initially localized quantum system with discrete energy spectrum \cite{r.robinett04a}, the discussion here may find its analogies in other physical systems, such as Rydberg atoms \cite{i.averbukh91a}, vibration of molecules \cite{i.averbukh91a}, or bosonic Josephson junction \cite{g.milburn97a}. Note that the faster beating dynamics of the uncertainty (or even higher moments) facilitate the detection of the quantum revivals, especially when the revival of the mean value is too long to be observed experimentally or is destroyed by decoherence.

Notice that the baseline of $\Delta \cos ^{2} \theta$, like that of $\langle \cos^{2}\theta \rangle$ \cite{p.ho09a}, is upshifted, with a strong-field limit $\Delta \cos ^{2} \theta  \rightarrow 1/\sqrt{8}$. Because the background of $\Delta \cos ^{2} \theta$ grows even higher than the value for an isotropic distribution, given the same maximum degree of alignment of $0.8$, molecules aligned impulsively have a fluctuation ($\approx 0.24$) higher than those aligned adiabatically ($\approx 0.17$).

The classical dynamics in Fig.~\ref{fig:impulsive} reproduce well the short-term quantum evolutions of both $\langle \cos^{2}\theta \rangle$ and $\Delta \cos ^{2} \theta$ until their breakdown first flagged by $\Delta \cos ^{2} \theta$ at $1/4\:\tau_{\text{rev}}$. Afterwards, the classical calculations show no revival but capture accurately the baselines of the quantum mean and uncertainty, which stem from the static population elements $\Delta J = 0$. It is interesting that the time average of the quantum mean coincides with the baseline level, so the classical treatment can predict the time averaged impulsive alignment process in reference to $\langle \cos^{2}\theta \rangle$ \cite{y.khodorkovsky11a, p.ho09a}. Yet, the time average of the quantum uncertainty is lower than the baseline, preserving the influence of the quantum coherence $\Delta J = \pm2$ due to the $\langle \cos^{2}\theta \rangle^2$ contribution.

Another way to inspect the dynamical information content of the uncertainty is to compare the time evolutions of $\langle \cos^{2}\theta \rangle$ and $\Delta \cos ^{2} \theta$ to that of the alignment distribution. Figures \ref{fig:impulsive}(c) and \ref{fig:impulsive}(d) plot the quantum and classical dynamics of the angular probability density distributions integrated over the azimuthal angle $\phi$, denoted by $\rho(\theta)$. The distributions indicate that the quantum and classical dynamics actually diverge much faster than indicated by $\langle \cos^{2}\theta \rangle$ and $\Delta \cos ^{2} \theta$. After this short period, the classical molecules disperse rapidly, leading to the enhanced backgrounds of $\langle \cos^{2}\theta \rangle$ and $\Delta \cos ^{2} \theta$. On the other hand, the quantum distribution soon develops a complex but regular interference pattern. The most outstanding features appear close to $1/2 \: \tau_{\text{rev}}$ and $\tau_{\text{rev}}$, as identified by $\langle \cos^{2}\theta \rangle$ and $\Delta \cos ^{2} \theta$. In the vicinity of $1/4 \tau_{\text{rev}}$ and $3/4 \tau_{\text{rev}}$ are the second salient features only captured by $\Delta \cos ^{2} \theta$. Thus, the additional dynamics of the uncertainty signifies the formation of an order of the alignment distribution beyond the description of the average order parameter. At other fractional revivals, various structures appear that are outside the grasp of $\langle \cos^{2}\theta \rangle$ and $\Delta \cos ^{2} \theta$ and are related to even higher moments of the distribution.

It is worth noting that it is possible to resolve the alignment distribution not only theoretically but also experimentally by Coulomb exploding the molecules and recording the angular-resolved momentum distribution of the ion fragments \cite{h.stapelfeldt03a, e.peronne04a, p.dooley03a}. Based on the measured distributions, both the average and the uncertainty of the alignment parameter can be constructed. Also, it is viable to find another complex observable whose expectation value is effectively sensitive to higher moments of $\cos^{2}\theta$, e.g., the high-harmonic-generation signals of impulsively aligned molecules \cite{s.ramakrishna07a, s.ramakrishna08a, a.abdurrouf09a, r.lock12a, s.weber13a}. Accordingly, the dynamics of the fluctuations of this specific system are measurable and have observable impacts on other measurements.

Even in a model system, the dynamics of fluctuations can be highly nontrivial. One should expect this to be even more important for a nonequilibrium quantum many-body system, where entanglement and correlations further participate in regulating fluctuations \cite{r.landauer98a}. To illustrate the basic idea, consider a system of $N$ interacting fermions at zero temperature with a normalized many-body wave function $\Ket{\Phi (t)}$ in the Schr\"{o}dinger picture \cite{a.szabo96a}. To proceed, it is instructive to introduce a series of reduced density matrices that is equivalent to a description in terms of the total wave function. The $p$-th order reduced density matrix ($p$-RDM) is defined by \cite{k.sakmann08a, p.lowdin55a}: $\rho^{(p)}(\mathbf{r}_1,\cdots, \mathbf{r}_p | \mathbf{r}'_1, \cdots, \mathbf{r}'_p;t)  \equiv  \Bra{\Phi} \hat{\Psi}^{\dagger}(\mathbf{r}'_1)  \cdots \hat{\Psi}^{\dagger}(\mathbf{r}'_p) \hat{\Psi}(\mathbf{r}_p) \cdots  \hat{\Psi}(\mathbf{r}_1) \Ket{\Phi}$, where $\hat{\Psi}(\mathbf{r}_i)$ is the fermionic field operator \cite{n.march95a} and $\mathbf{r}_i$ a vector of spatial coordinates (for simplicity, we suppress the spin). For a general local one-body observable $\hat{O}$, we have:
\begin{align}
\langle \hat{O} \rangle (t) &= \int d^3r_1 \left\{ \hat{O}_{\mathbf{r}_1} \rho^{(1)}(\mathbf{r}_1|\mathbf{r}'_1;t) \right\}_{\mathbf{r}'_1 = \mathbf{r}_1}  \label{eq:average} \\
( \Delta O )^2(t) &= \iint d^3 r_1 d^3 r_2 \left\{ \hat{O}_{\mathbf{r}_1} \hat{O}_{\mathbf{r}_2} \right. \nonumber \\
& \left( \rho^{(2)} (\mathbf{r}_1,\mathbf{r}_2|\mathbf{r}'_1,\mathbf{r}'_2 ; t) + \rho^{(1)} (\mathbf{r}_1|\mathbf{r}'_1 ; t) \delta (\mathbf{r}_1-\mathbf{r}_2) \right.   \nonumber \\
& \left. \left. -  \rho^{(1)}(\mathbf{r}_1|\mathbf{r}'_1 ; t) \rho^{(1)}(\mathbf{r}_2|\mathbf{r}'_2 ; t) \right) \right\}_{ \substack{\mathbf{r}'_1 = \mathbf{r}_1 \\ \mathbf{r}'_2 = \mathbf{r}_2}},\label{eq:uncertainty}
\end{align}
where $\hat{O}_{\mathbf{r}_i}$ stands for an operator acting only on the unprimed coordinate $\mathbf{r}_i$ \cite{p.lowdin55a}. As the $p$-RDM is identical to the $p$-th order spatial correlation function \cite{k.sakmann08a, r.glauber63a}, it is evident that the information mapped out independently by $\Delta O$ is the 2-RDM, which directly measures two-particle properties and has diagonal terms (i.e.~$\mathbf{r}'_i = \mathbf{r}_i$) being the pair density \cite{p.lowdin55a}. The functional form of the 2-RDM is very sensitive to entanglement and correlations: For classical noninteracting particles, the 2-RDM can be perfectly factorized; for noninteracting fermions whose wave function is a single Slater determinant, the 2-RDM can be reduced to antisymmetrized products of 1-RDMs \cite{, p.lowdin55a, n.march95a}. On the other hand, $\langle \hat{O} \rangle$ merely reveals information on the 1-RDM, which measures one-particle properties and has diagonal terms being the particle density \cite{p.lowdin55a}. Hence, analyzing the dynamics of the uncertainty of a one-body observable provides a general approach to access information on the time evolution of the second-order correlation function. If $\hat{O}_{\mathbf{r}_i}$ does not contain any derivative with respect to the spatial coordinates, i.e., $\hat{O}_{\mathbf{r}_i} = O({\mathbf{r}_i})$, then only the diagonal elements of RDMs are needed \cite{p.lowdin55a} and the two terms in the second line of Eq.~(\ref{eq:uncertainty}) become the density-density correlation function \cite{m.naraschewski99a}. It is possible to directly construct the density-density correlation function from a measurement of a set of one-body observables. For example, we mention the impressive works on BEC utilizing absorption images \cite{e.altman04a, m.greiner05a, s.folling05a} or single-atom-resolved \textit{in situ} imaging technique \cite{j.sherson10a, w.bakr10a}, very recently even to probe the time-resolved density-density correlation function \cite{m.cheneau12a, p.schaus12a, t.fukuhara13a}. Yet, it is generally more feasible in other cases to extract condensed information on the density-density correlation function or less restrictedly the 2-RDM based on the noise analysis of a measurement of a single one-body observable.

To conclude, using laser-aligned molecules as a concrete example of a highly nonequilibrium quantum system, we demonstrate that the uncertainty of the alignment order parameter can probe nontrivial and intricate dynamics remarkably distinct from those shown by the average alignment order parameter, bringing information inaccessible by the average. Also, we illustrate that the dynamics of the uncertainty are more sensitive to the discrepancy between quantum and classical approaches. Lastly, we remark that the dynamics of the uncertainty of a one-body observable offer a general and valuable tool for studying real-time two-body correlations in a many-body system, due to its close connection with a two-body operator and the second-order correlation function. For more than a century, fluctuations have kept revolutionizing the way we understand fundamental questions ranging from particle-wave duality \cite{a.einstein09a, j.johnson28a, h.nyquist28a} to entanglement and correlations \cite{r.hanburybrown56a, r.landauer98a, c.beenakker03a}. With the presented model system as proof of principle and with the rapid advancement of technology, opportunities are envisioned for the understanding of nonequilibrium dynamics of quantum systems founded on the dynamics of fluctuations.

\begin{acknowledgments}
Y.-J.C.~thanks M.~Eckstein, Y.~Laplace, and N.~Rohringer for stimulating input from various fields of research, and S.-K.~Son for carefully reading the manuscript.
\end{acknowledgments}

%

\end{document}